# NASICON solid-electrolyte modification and analysis using ion and neutron beams

Giovanni Ceccio[1,*], Jiri Vacík[1], Mykhailo Drozdenko[1], Romana Mikšová[1], Ivan Mastronardo[2,3], Dejan Prokop[4,5], Benedetta Brancato[2,3], Eva Štěpanovská[1], Claudia D'Urso[2], Leone Frusteri[2]

[1]Department of Neutron and Ion Methods, Nuclear Physics Institute (NPI) of the Czech Academy of Sciences, 25068 Husinec-Řež, Czech Republic
[2]CNR-ITAE, Via Santa Lucia sopra contesse, 5 – 98121, Messina (ME), Italy
[3]University of Messina. Department of Engineering, Contrada di Dio, 98158 Messina (ME), Italy
[4]Charles University, Faculty of Mathematics and Physics, V Holešovičkách 747/2, 180 00 Prague, Czech Republic
[5] Institute of Physics of the Czech Academy of Sciences, Na Slovance 1999/2, 182 00 Prague, Czech Republic

Corresponding author: ceccio@ujf.cas.cz

**Abstract**

Solid electrolytes (SEs) for sodium-based superionic conductors (NaSICON) are widely recognized for their excellent ionic conductivity and application in sodium based energy storage systems. While considerable effort has been made to develop thin electrolytes for all-solid-state batteries (ASSBs) for lithium ions, only a few sodium-based SEs have been successfully fabricated as thin films. These thin films are particularly desirable for their reduced electrical resistance, which typically increases with the thickness of the SE. By reducing the thickness of the SEs to the nanometer scale, their ionic conductivity can be significantly enhanced. In this study, the NASICON composite was initially prepared in the form of pellets using the mixed oxide technique with a planetary ball mill and synthesized by the solid-state method at 1250 °C. The resulting pellets were used as sputtering targets in a low-energy ion facility to prepare continuous and uniform NASICON nanofilms. To explore the effect of ion implantation on the electrical properties of NASICON, the prepared films were bombarded with Ni ions at 1.1 MeV and varying fluences, using the Tandetron accelerator at the CANAM infrastructure (NPI Řež). The electrical properties of both the synthesized and implanted films were analyzed through electrochemical impedance spectroscopy (EIS). The results, describing the impact of irradiation on NASICON's properties, are presented here.

## 1. Introduction

Growing environmental concerns and the increasing deficiency of energy resources have become critical global issues due to the depletion of fossil fuel reserves. Renewable energy sources such as wind, solar, and tidal power are inherently random, intermittent, and unstable, whereas electricity consumption follows distinct temporal and spatial patterns. This mismatch has intensified the need for energy storage systems that can ensure long term stability and reliability. Among various storage technologies, lithium-ion batteries (LIBs) currently dominate the market worldwide due to their high energy density, excellent coulombic efficiency, and low self-discharge rate [1, 2, 3]. They are extensively utilized in electric vehicles (EVs), aerospace applications, portable electronics, and numerous other fields [4, 5, 6]. Nevertheless, the escalating cost and limited availability of lithium pose significant challenges to the large-scale deployment of LIBs in future energy infrastructures. This has accelerated the search for alternative energy storage technologies. Sodium metal with its low redox potential, high specific capacity and high abundance, attracted attention as possible anode material for implementation of sodium batteries [7, 8, 9, 10]. Like the lithium counterpart, the risk of liquid electrolyte leak with sub-sequential safety risk, was a promoter of the will to search for an "All solid state" solution, where a solid-state-electrolyte is needed [11, 12]. Such SSE must possess good mechanical and electrical properties, acting as physical barrier for dendrites formation and allowing the cycling of the Na ions. Solid electrolytes (SEs) for sodium-based superionic conductors (NaSICON) were first introduced in 1976 and quickly recognized for their excellent ionic conductivity [13, 14, 15]. Recent research has identified NZSP electrolytes as promising candidates for sodium-ion rechargeable batteries in low-temperature/room-temperature applications, owing to their superior ionic conductivity. A further step for wider spread of such system lies in the possibility to use it in ultrathin all solid state systems for nano-thick batteries. In this work we performed a pioneer investigation for the possibility to form nanofilms of NASICON and for their modification using ion beam implantation, in order to act directly on the electrochemical properties by implantation of ferromagnetic materials.

## 2. Experimental section

### 2.1 Materials

For the preparation of NASICON pellets, a solid state reaction method was used. NZSP was synthesized by mixing Ammonium Phosphate monobasic (Sigma Aldrich), Zirconium Dioxide (BDH Chemicals), Sodium Carbonate (Honeywell Fluka) and Silicon Dioxide fumed (Sigma Aldrich) in stoichiometric ration to produce $Na_3Zr_2Si_2PO_{12}$. The precursors were ball-milled in a planetary system with a Zirconia pot. This mixture was then pressed in pellet and calcined at 1150 °C for 4 hours with ramp up of 300 °C/h. After this first heating treatment the pellet was then ball-milled again and reshaped in pellet for a final sintering at 1250°C with the same time and ramp up. A reference Pellet (Namely Pellet 2) was synthesized with the same method starting from commercial NZSP powder (MSE Supplies). The labmade NaSICON was used to prepare two different pellets. For the NZSP-4 the standard procedure described earlier was used but for the pellet NZSP-5 the second ball milling step was skipped.

### 2.2 NASICON deposition

Thin films of NASICON were then produced from these pellets using the ion beam sputtering (IBS) technique, with the pellets serving as targets for the sputtering. The ion beam sputtering was performed at the LEIF facility in NPI Řež. LEIF (Low Energy Ion Facility) is a multipurpose laboratory-made ion beam system that generates high intense ion beam from a duoplasmatron ion source and operates within the CANAM infrastructure. Ion beam energy may range between 1 keV and 100 keV with a beam current that can reach several mA. In order to perform sputtering of the NaSICON pellets, Ar ion beam was used at 20 kV and focused into a multipurpose vacuum chamber, where it irradiates the NaSICON pellet at a 45° angle, sputtering the pellet material. Silicon chips <100> (Tedpella, Inc.), placed looking at the target with an inclination of 30° (out of the beam's direct path in a distance of about 15 cm), served as substrates. During the IBS, the substrates were kept at room temperature. The produced films were named after the original pellets as NASICON 2, NASICON 4 and NASICON 5.

## 2.3 NASICON modification

The thin films were modified by ion technologies in order to change electric properties of produced films. Ion implantation is a well known technique for the modification in deep of material, changing the surface properties by doping with selected atoms. Ion implantation was performed using 1.1 MeV Ni ions at three different fluences, 1E14 $Ni/cm^2$, 5E14 $Ni/cm^2$ and 1E15 $Ni/cm^2$. At the used energies the ions will pass through the coating layer entirely, interacting primarily with the substrate. The film properties are modified via electronic stopping, due to the ion's energy deposited into the electronic system of the films, inducing ionization, bond disruption, and structural changes without significant atomic displacement.

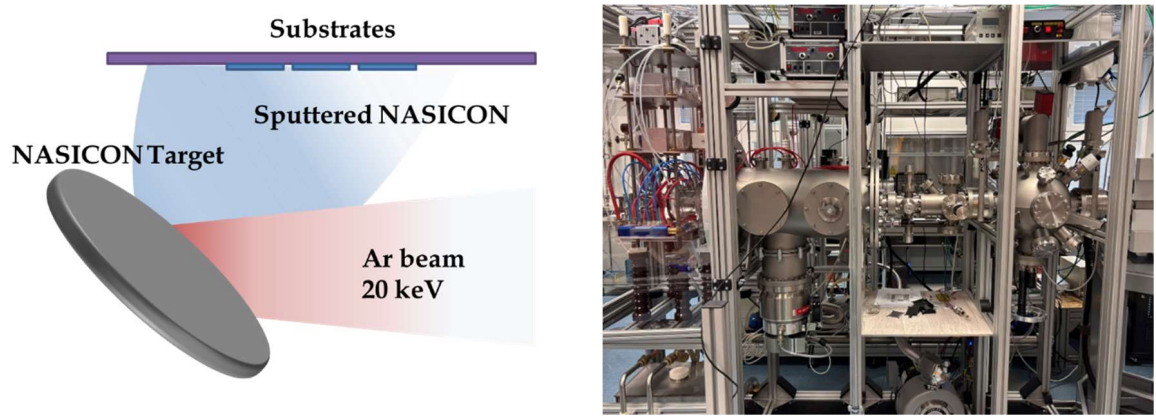

**Fig. 1.** Sketch of the IBS of NaSICON a), picture of the LEIF facility b).

## 2.4 Characterization

A comprehensive set of characterization techniques was employed to investigate the pellets and the produced thin films. Surface morphologies of samples were observed by SEM (TM4000 Plus, Hitachi High-Tech Europe GmbH, Germany). Selected specimens were mounted on specimen stubs by double-sided adhesive carbon tape, placed into an electron microscope holder and observed using backscattered electron mode (BSE), at an accelerating voltage of 15 kV. The XRD diffraction measurements were performed by Bruker D8 DISCOVER diffractometer, equipped with a four-crystal Bartels monochromator and Cu $K_\alpha$ lamp, in Bragg-Brentano asymmetric 2θ-ω scan (to exclude contribution corresponding to diffraction from Si) in range from 5 degrees to 115 degree. The ionic conductivity at room temperature was evaluated based on the electrochemical impedance spectroscopy (EIS) measurement. To perform EIS, Biologic VSP-300 Potentiostat was used, operating in frequency range 7 MHz – 0.1Hz with a perturbation of 20mV. All measurements were recorded at room temperature (25°C) using a two-electrode setup.

**Results**

Fig, 2 shows the micrograph obtained for the pristine and irradiated samples. In particular, it is possible to see all samples pristine show a flat and uniform surface with a minimum amount of the surface structures. The films produced from the commercial NaSICON after irradiation presented formation of nanostructures on the surface for all irradiation doses, with a slightly higher number of structures for the higher dose. For the lab-made NaSICON, the relatively produced films show less structure production, the films produced by NaSICON 4 show formation of big clusterization at higher dose, meanwhile the NaSICON 5 shows homogeneous films after all irradiation.

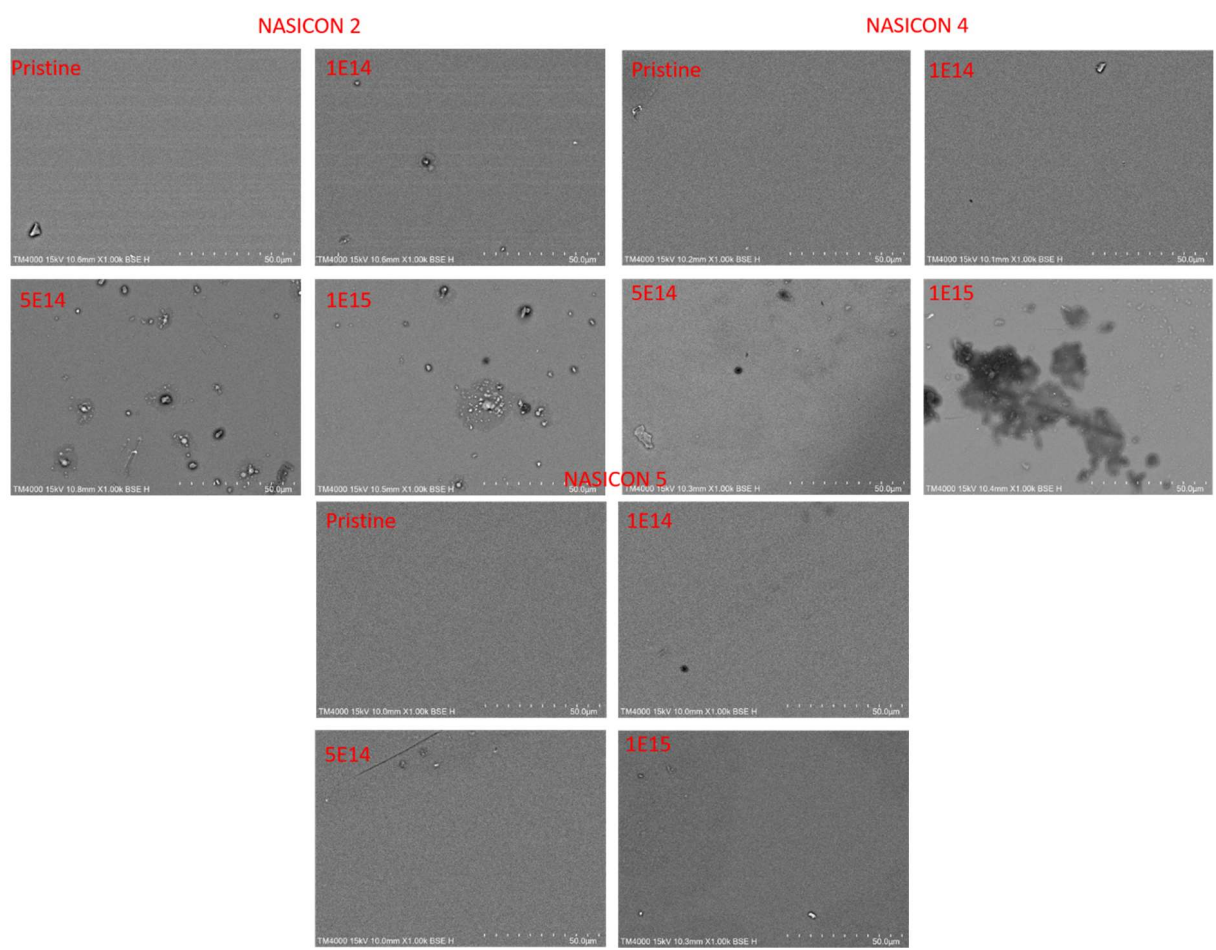

**Fig. 2.** SEM micrograph for pristine and irradiated films.

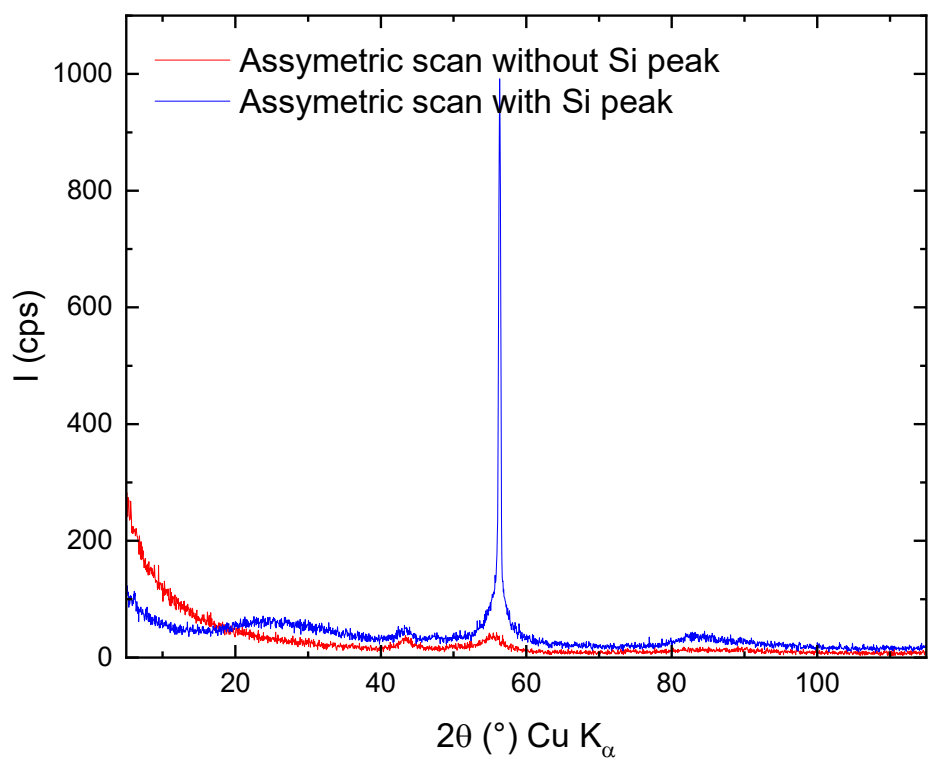

**Fig. 3.** Diffractogram of the deposited film.

XRD results are shown in Fig.3 where it is possible to see the diffractogram with the presence of Si peak from the crystal substrate of Si (blue in figure) and the diffractogram with the removed mentioned peak. Peaks that could be associated with crystalline planes of the NaSICON are not present. In the blue diffractogram is possible to notice the presence of a small broad peak in the range between 18 and 39 2θ that can be linked to the presence of amorphous NaSICON [16, 17]. Generally, the NaSICON structure is induced by high temperature sintering, meanwhile our deposition was performed at room temperature. When NaSICON thin films are in this amorphous state, the material fundamentally lacks the long-range crystal order that defines the NaSICON framework. In practice, this means that instead of well-defined channels for Na ions migration, the structure becomes a disordered network where ion transport occurs by hopping between locally favorable sites. Should be expected also different density compared with crystal NaSICON, in particular the films can have more structural "free space," which influences diffusivity. Because the film is amorphous, the movement of Na ions cannot follow any defined crystallographic channels as it would in crystalline NaSICON. Instead, ion transport occurs through a disordered network, where ions hop between locally available sites. As a result, electrochemical impedance spectroscopy (EIS) may show a higher overall resistance and a higher activation energy for ion conduction.

From all the Nyquist plots (Fig. 4) it is possible to see a noticeable effect of the Ni implantation on the electrochemical properties of the NZSP films. For all the samples the treatment with the lowest Ni fluence induces an increase in the impedance in the material. This could be due to the initial formation of structural damage that disrupts the ionic transport pathways in the amorphous NZSP.

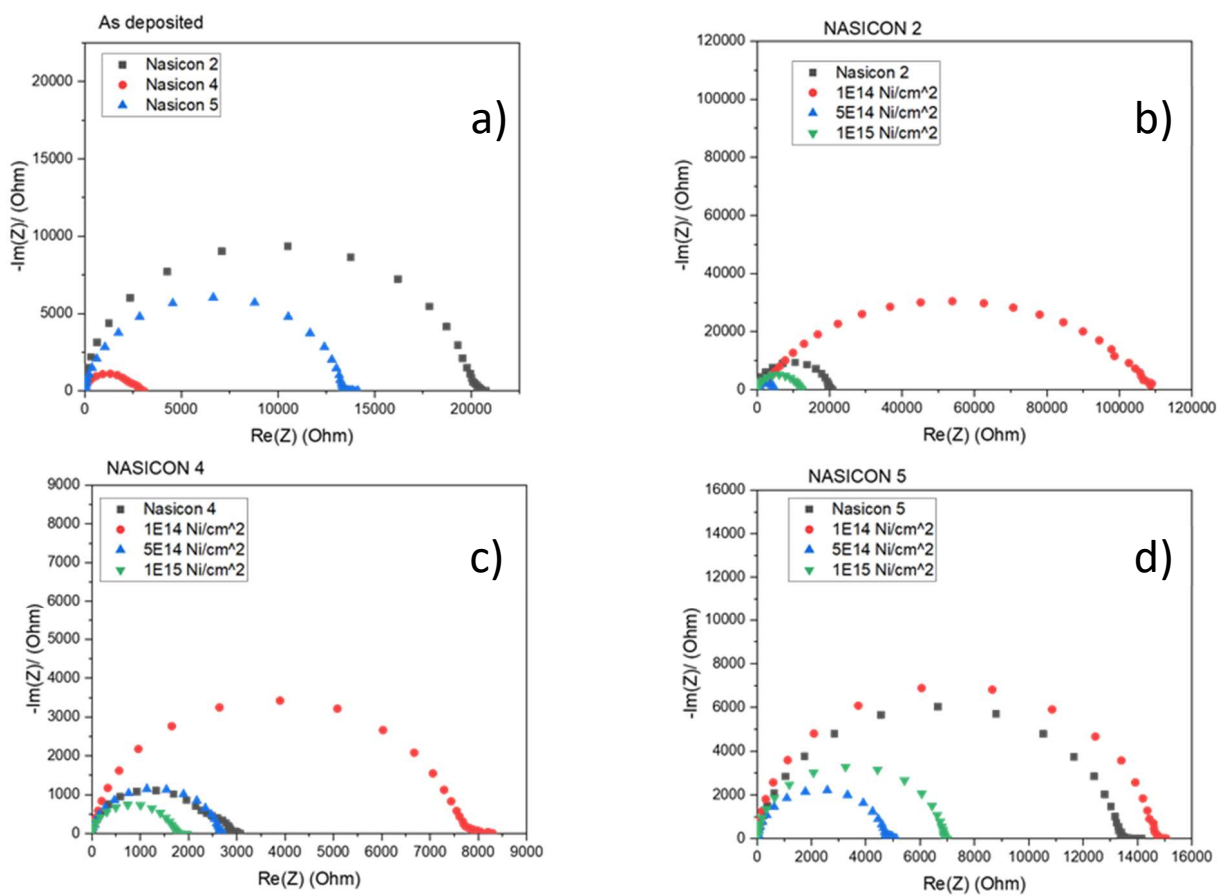

**Fig. 4.** Electrochemical Impedance Spectroscopy of NaSICON films, deposited and irradiated. a) comparison between al deposited film before irradiation. b), c) and d) shows the NASICON 2, 4, 5 respectively, as deposited and after irradiation.

But with the increase of the fluence it is possible to see a reduction in the impedance of the material suggesting an improvement of the ionic conductivity of the sample. The increased density of structural damages and vacancies generated inside the thin film at higher fluences could lead to the formation of vacant sites that inside a poorly crystalline material leads to improved conduction. This non-linear behavior of the material suggests that there could be an optimal amount of $Ni/cm^2$ for except for the NASICON 4 which shows a slightly lower one for the $1E1015\ Ni/cm^2$ this type of materials. Most of the samples show a minimum in impedance at $5E1014\ Ni/cm^2$.

## Conclusions

In this work, a pioneering investigation was carried out on the fabrication and modification of ultrathin NaSICON solid electrolyte films using ion beam based techniques, with the aim of exploring their potential for application in sodium-ion all-solid-state batteries. Bulk NaSICON pellets were successfully synthesized via a conventional solid-state route and subsequently employed as sputtering targets for ion beam sputtering, enabling the deposition of continuous and uniform NASICON nanofilms on substrates. Structural characterization revealed that the deposited films are predominantly amorphous, a consequence of the low-temperature deposition conditions, which suppress the formation of the long-range crystalline NaSICON framework typically obtained after high-temperature sintering.

Despite the amorphous nature of the films, with higher overall resistance and a higher activation energy for ion conduction, the electrochemical impedance spectroscopy demonstrated that ion beam modification plays a significant role in tuning the electrical properties of NaSICON thin films. Nickel ion implantation at 1.1 MeV induced pronounced, fluence-dependent changes in impedance behavior. At low implantation fluence, an increase in impedance was observed, likely associated with initial structural disorder and disruption of existing ionic transport pathways. Conversely, higher fluences led to a reduction in impedance for most samples, suggesting enhanced ionic conductivity. This improvement can be attributed to the accumulation of irradiation-induced defects, such as vacancies and locally disordered regions, which may facilitate sodium-ion hopping in an amorphous matrix. The observed non-linear response indicates the existence of an optimal implantation fluence, identified around 5E14 Ni/cm$^2$ for most films, beyond which further damage may counteract the beneficial effects. Overall, these results demonstrate that low-energy ion beam sputtering combined with controlled ion implantation represents a promising strategy for the fabrication and functional tuning of ultrathin NaSICON electrolytes. This approach opens new perspectives for engineering solid electrolytes at the nanoscale, where defect chemistry and disorder can be deliberately exploited to enhance ionic transport. Future studies will focus on: studying the eventual development of crystallinity of the nanofilm trough annealing treatment during deposition and post deposition, in addition to correlating implantation-induced structural changes with transport mechanisms and extending this methodology to other dopant species and solid electrolyte systems for next-generation sodium-based energy storage devices.

## Acknowledgements

This work was supported by the Ministry of Education, Youth and Sports (MEYS) CR under the project OP JAK CZ.02.01.01/00/22_008/0004591. The authors also acknowledged the support of the Czech Academy of Science Mobility Plus Project, Grant No. CNR-25-01.